\documentclass[12pt]{article}
\usepackage{amsfonts, amsbsy}
\DeclareMathAlphabet{\bi}{OML}{cmm}{b}{i}
\newcommand{\fl}{}
\newcommand{\etal}{\textit{et al.}}
\newcommand{\be}{\begin{equation}}
\newcommand{\ee}{\end{equation}}
\newcommand{\ba}{\begin{array}}
\newcommand{\ea}{\end{array}}
\newcommand{\p}{\partial}

\newcommand{\ds}{\displaystyle}

\newtheorem{prop}{Proposition}

\begin{document}

\title
{\vspace*{-1cm}A strange recursion operator demystified}

\author{A. Sergyeyev\\ %
Silesian University in Opava, Mathematical Institute,\\ Na Rybn\'\i{}\v{c}ku 1,
746\,01 Opava, Czech Republic\\
E-mail: {\tt Artur.Sergyeyev@math.slu.cz}}
\date{}
\maketitle

\begin{abstract}
We show that a new integrable two-component system of KdV type
studied by
Karasu (Kalkanl\i{}) et al.\
(arXiv: {\tt nlin.SI/0203036}) is bihamiltonian,
and its recursion operator, which has a highly unusual structure
of nonlocal terms,
can be written as a ratio of two compatible Hamiltonian operators found by us.
Using this we prove that the system in question possesses an
infinite hierarchy of {\em local} commuting generalized
symmetries and conserved quantities in involution,
and the evolution systems corresponding to these
symmetries are bihamiltonian as well.
\looseness=-1

We also show that upon introduction of suitable nonlocal
variables the nonlocal terms of the recursion operator under study
can be written in the usual form, with the integration operator
$D^{-1}$ appearing in each term at most once.
\end{abstract}

%



Using the Panilev\'e test, Karasu (Kalkanl\i) \cite{K97} and Sakovich \cite{S99}
found a new integrable evolution system of KdV type,
\be \ba{l}
u_{t}=4 u_{xxx}-v_{xxx}-12 u u_x+v u_x+2 u v_x,\nonumber\\
v_{t}=9 u_{xxx}-2 v_{xxx}-12 v u_x-6 u v_x+4 v v_x, \label{eveq}
\ea
\ee
and a zero curvature representation for it \cite{S99}.
Notice that this system is, up to a linear transformation of $u$ and $v$,
equivalent to the system (16) from the Foursov's \cite{four} list of
two-component evolution systems of KdV type possessing (homogeneous)
symmetries of order $k$, $4\leq k\leq 9$.

Karasu (Kalkanl\i), Karasu and Sakovich \cite{ka}
found that (\ref{eveq}) has a recursion operator of the form
$$
\ba{l}
\mathfrak{R}=\left(
\ba{cc}
R_{11} & R_{12}\\
R_{21} & R_{22}%
\ea\right),
\ea
$$
where
\[
\ba{l}
R_{11}=3D_{x}^{2}-6u-3u_{x}D_{x}^{-1}, \\[2mm]
R_{12}=\left[  -2D_{x}^{5}+\left(  2u+3v\right)
D_{x}^{3}+\left(
8v_{x}-4u_{x}\right)  D_{x}^{2} \right. 
 +\left(7v_{xx}-6u_{xx}+4u^{2}-6uv\right)  D_{x}\nonumber\\[2mm]
 -2u_{xxx}+2v_{xxx}+6uu_{x}-3vu_{x}-4uv_{x}
  \left.  +u_{x}D_{x}^{-1}\circ v_{x}\right] \circ \left(  3D_{x}^{3}-4vD_{x}%
-2v_{x}\right)^{-1}, \\[2mm]
R_{21}=6D_{x}^{2}+6u-9v-3v_{x}D_{x}^{-1}, \\[2mm]
R_{22}  =\left[  -3D_{x}^{5}+\left(12v-18u\right)
D_{x}^{3}+\left(18v_{x}-27u_{x}\right)  D_{x}^{2}\right. 
 +\left(14v_{xx}-21u_{xx}+12u^{2}\right. 
 \left.+12uv-9v^{2}\right)  D_{x} \nonumber\\[2mm]
-6u_{xxx}+4v_{xxx}+12uu_{x}+6vu_{x}+6uv_{x}-9vv_{x}
 \left.  +v_{x}D_{x}^{-1}\circ v_{x}\right] \circ \left(3 D_{x}^{3}-4vD_{x}%
-2v_{x}\right)^{-1}.
\ea
\]

Here $D_x$ is the operator of total $x$-derivative:
$$
D_x=\p/\p x+u_x\p/\p u+v_x\p/\p v
+\sum_{j=2}^{\infty} (u_{jx}\p/\p u_{(j-1)x}+ v_{jx}\p/\p v_{(j-1)x}),
$$
$u_{kx}=\p^k u/\p x^k$, and $v_{kx}=\p^k v/\p x^k$, see e.g.\ \cite{olv_eng2}
for further details.

Let also
$$
\delta/\delta u=\p/\p u+\sum_{j=1}^{\infty}(-D_x)^j\p/\p u_{jx},\quad
\delta/\delta v=\p/\p v +\sum_{j=1}^{\infty}(-D_x)^j\p/\p v_{jx}.
$$
We shall use the notation $\bi{u}=(u,v)^T$,
and $\delta/\delta\bi{u}=(\delta/\delta u,\delta/\delta v)^T$ will
stand for the usual operator of variational derivative,
cf.\ e.g.\ \cite{olv_eng2}.
Here and below the superscript $T$ denotes the matrix transposition.
Recall that a function that depends on $x,t,u,v$ and a {\em finite}
number of $u_{jx}$ and $v_{kx}$ is said to be {\em local}, see e.g.~\cite{olv_eng2, mik1}.
\looseness=-1

Because of the nonstandard structure of nonlocal terms
in $\mathfrak{R}$ the known `direct' methods
(see e.g.\ \cite{sw, serg1, serg2, serg3} and references therein)
for proving the locality of symmetries generated by $\mathfrak{R}$
are not applicable, so the question of whether (\ref{eveq})
has an infinite hierarchy of local commuting symmetries
remained open for a while.
It also was unknown
whether (\ref{eveq}) is a bihamiltonian
system.



We have \cite{ka}
$\mathfrak{R}=\mathfrak{M}\circ \mathfrak{N}^{-1}$, where $\mathfrak{M}$ and $\mathfrak{N}$ are some
(non-Hamiltonian) differential operators of order five and three.
%
Inspired by this fact, we undertook a search of Hamiltonian operators
of order three and five for (\ref{eveq}), and it turned out that such operators do exist
and (\ref{eveq}) is bihamiltonian.
Namely, the following assertion holds.
\looseness=-1
\begin{prop}\label{propbih}The system (\ref{eveq}) is bihamiltonian:
\be\label{bih}
\bi{u}_t=\mathfrak{P}_1\delta H_0/\delta\bi{u}=\mathfrak{P}_0\delta H_1/\delta\bi{u},
\ee
where $H_0=-3 u+v/2$, $H_1=2 u^2- u v+v^2/9$,
and $\mathfrak{P}_0$ and $\mathfrak{P}_1$ are compatible Hamiltonian operators of the form
\be\label{p}
\fl\mathfrak{P}_0=\left(
\ba{cc}
D_x^3- 2 u D_x -u_x & 0\\
0 & -9 D_x^3+12 v D_x +6 v_x %
\ea\right),
\mathfrak{P}_1=\left(
\ba{cc}
P_{11} & P_{12}\\
P_{21} & P_{22}%
\ea\right),
\ee
where
$$
\ba{l}
P_{11}=D_x^5- 4 u D_x^3 - 6 u_x D_x^2 +4(u^2-u_{xx}) D_x -u_{xxx}+ 4 u u_x-u_x D_x^{-1}\circ u_x,\\
P_{12}=2 D_x^5- (2 u+3v) D_x^3+ 4 (u_x - 2 v_x)D_x^2 +(6 u_{xx}-7 v_{xx}-4 u^2+6 u v) D_x\\
+2 u_{xxx}-2 v_{xxx}-6 u u_x+ 3 v u_x+ 4 u v_x-u_x D_x^{-1}\circ v_x,\\
P_{21}=2 D_x^5- (2 u+3v) D_x^3 - (10 u_x+v_x) D_x^2 +(-4 u^2+6 u v- 8 u_{xx}) D_x\\
- 2 u_{xxx} -2 u u_x+ 3v u_x+2 u v_x-v_x D_x^{-1}\circ u_x,\\
P_{22}=3 D_x^5+ (18 u -12 v)D_x^3 + (27 u_x -12 v_x)D_x^2 +(21 u_{xx}-14 v_{xx}-12 u^2-12 u v+9 v^2) D_x\\
+6 u_{xxx}-4 v_{xxx}-12 u u_x-6v u_x-6 u v_x+9 v v_x-v_x D_x^{-1}\circ v_x.
\ea
$$

Moreover, we have $\mathfrak{R}=3 \mathfrak{P}_1\circ\mathfrak{P}_0^{-1}$,
and hence $\mathfrak{R}$ is hereditary.
\end{prop}
Now we are ready to prove that (\ref{eveq}) has infinitely many local commuting symmetries.
\begin{prop}\label{bihloc}
%
Define the quantities $\bi{Q}_{j}$ and $H_j$ recursively by the formula
$\bi{Q}_{j}=\mathfrak{P}_1\delta H_{j}/\delta\bi{u}=\mathfrak{P}_0 \delta H_{j+1}/\delta\bi{u}$,
$j=0,1,2,\dots$, where
$H_0$, $H_1$, $\mathfrak{P}_0$ and $\mathfrak{P}_1$ are given in Proposition~\ref{propbih}.
Then $H_j$, $j=2,3,\dots$, are local functions that can be chosen to be independent of $x$ and $t$,
and $\bi{Q}_{j}$ are local
commuting generalized symmetries for (\ref{eveq}) for all $j=1,2,\dots$.
\looseness=-1

Moreover, the evolution systems $\bi{u}_{t_j}=\bi{Q}_j$ are bihamiltonian with respect to
$\mathfrak{P}_1$ and $\mathfrak{P}_0$ by construction,
and $\mathcal{H}_j=\int H_j dx$
are in involution with respect to the Poisson brackets associated
with $\mathfrak{P}_0$ and $\mathfrak{P}_1$
for all $j=0,1,2\dots$, so $\mathcal{H}_j$ are
common conserved quantities for all evolution systems $\bi{u}_{t_k}=\bi{Q}_k$, $k=0,1,2,\dots$.
\end{prop}

{\em Proof.}
%
%
Let us use induction on $j$. Assume that 
for $\bi{Q}_{j}=\mathfrak{P}_1\delta H_{j}/\delta\bi{u}$
there exist a local function $H_{j+1}$
such that $\bi{Q}_{j}=\mathfrak{P}_0 \delta H_{j+1}/\delta\bi{u}$
and $\p H_{j+1}/\p x=\p H_{j+1}/\p t=0$,
and let us show that then $\bi{Q}_{j+1}=\mathfrak{P}_1\delta H_{j+1}/\delta\bi{u}$ is local too and
there exists a local function $H_{j+2}$ such that
$\bi{Q}_{j+1}=\mathfrak{P}_0 \delta H_{j+2}/\delta\bi{u}$ and $\p H_{j+2}/\p x=\p H_{j+2}/\p t=0$.

Since $H_j$ is independent of $x$, we have
\be\label{loc}
u_x \frac{\delta H_j}{\delta u}
+v_x\frac{\delta H_j}{\delta v}
=D_x\biggl(H_j-\sum_{j=1}^{\infty}\sum_{k=0}^{j-1}
\biggl\{u_{(j-k)x}(-D_x)^k\biggl(\frac{\p H}{\p u_{jx}}\biggr)
+v_{(j-k)x}(-D_x)^k\biggl(\frac{\p H}{\p v_{jx}}\biggr)\biggr\}\biggr).
\ee
As $H_j$ is local by assumption, the sum is actually finite, so
$D_x^{-1}(u_x \delta H_j/\delta u+v_x \delta H_j/\delta v)$
is a local expression, and hence
$\bi{Q}_{j+1}=\mathfrak{P}_1\delta H_j/\delta\bi{u}$ is local too.

Next, as $\bi{Q}_{j+1}=\mathfrak{P}_1\delta H_{j+1}/\delta\bi{u}$,
$\bi{Q}_{j}=\mathfrak{P}_0 \delta H_{j+1}/\delta\bi{u}$, and $\mathfrak{R}=
3\mathfrak{P}_1\circ \mathfrak{P}_0^{-1}$, we can (formally) write
$\bi{Q}_{j+1}=(1/3)\mathfrak{R}\bi{Q}_j$, cf.\ e.g.\ Section 7.3 of \cite{olv_eng2}.
As $\mathfrak{R}$ is a recursion operator for (\ref{eveq}), its
Lie derivative along $\bi{Q}_0$ vanishes:
$L_{\bi{Q}_0}(\mathfrak{R})=0$. By Proposition~\ref{propbih}
$\mathfrak{R}$ is hereditary, so we have \cite{ff}
$L_{\bi{Q}_{j+1}}(\mathfrak{R})=(1/3)^{j+1}L_{\mathfrak{R}^{j+1}\bi{Q}_0}(\mathfrak{R})=0$,
whence $L_{\bi{Q}_{j+1}}(\mathfrak{P}_0)=3 L_{\bi{Q}_{j+1}}(\mathfrak{R}^{-1}\circ\mathfrak{P}_1)=
3\mathfrak{R}^{-1}\circ L_{\bi{Q}_{j+1}}(\mathfrak{P}_1)
=3\mathfrak{R}^{-1}\circ L_{\mathfrak{P}_1\delta H_{j+1}/\delta\bi{u}}
(\mathfrak{P}_1)=0$, cf.\ e.g.\ \cite{bl}.
\looseness=-1

In turn, $L_{\bi{Q}_{j+1}}(\mathfrak{P}_0)=0$
implies that there exists a local function $H_{j+2}$ such that
$\bi{Q}_{j+1}=\mathfrak{P}_0\delta H_{j+2}/\delta\bi{u}$.
This is proved
along the same lines as 
in~\cite{olvbih}
for the second Hamiltonian structure of the KdV
equation.
Finally, as the coefficients of
$\mathfrak{P}_0$ and $\mathfrak{P}_1$ are independent of $x$ and $t$,
it is immediate that we always
can choose $H_{j+2}$ so
that it is independent of $x$ and $t$.
\looseness=-1

The induction on $j$ starting from $j=0$
and the application of Theorem 7.24 from Olver \cite{olv_eng2}
complete the proof. $\square$

For any local $H$
such that $\p H/\p x=0$ we shall set, in
agreement with (\ref{loc}) 
(see e.g.\ \cite{guthrie}--\cite{swro} for
more details on dealing with nonlocalities),
$$
D_x^{-1}\left(u_x \frac{\delta H_j}{\delta u}
+v_x\frac{\delta H_j}{\delta v}\right)
=H_j-\sum_{j=1}^{\infty}\sum_{k=0}^{j-1}
\left\{u_{(j-k)x}(-D_x)^k\left(\frac{\p H}{\p u_{jx}}\right)
+v_{(j-k)x}(-D_x)^k\left(\frac{\p H}{\p v_{jx}}\right)\right\}.
$$

Then, for instance, the first commuting flow for (\ref{eveq}) reads
$$
\ba{l}
u_{t_1}=2 u_{5x}-(5/9)v_{5x}-20 u u_{xxx}+(50/9) u v_{xxx}+(40/9)v u_{xxx}-(10/9)v v_{xxx}\\[2mm]
-50 u_x u_{xx}+(125/9) u_x v_{xx}+(40/3) v_x u_{xx}-(10/3)v_x v_{xx}
-(40/3)v u u_x+(20/9) v u v_x\\[2mm]
+40 u^2 u_x-(80/9) u^2 v_x+(5/9) v^2 u_x,\\[2mm]
v_{t_1}=5 u_{5x}-(4/3)v_{5x}-40 u u_{xxx}+10 u v_{xxx}+(10/3) v u_{xxx}-(5/9) v v_{xxx}
-120 u_x u_{xx}\\[2mm]
+30 u_x v_{xx}+(80/3)v_x u_{xx}-(55/9)v_x v_{xx}
+(160/3) v u u_x-20 v u v_x
+(40/3) u^2 v_x\\[2mm]
-(40/3)v^2 u_x+(35/9)v^2 v_x.
\ea
$$

By Proposition~\ref{bihloc} this system is bihamiltonian, and indeed we can write it as
$$
u_{t_1}=\mathfrak{P}_1\delta H_1/\delta\bi{u}=\mathfrak{P}_0\delta H_2/\delta\bi{u},
$$
where $$
H_2={\ds\frac{7}{162}}v^3-{\ds\frac83} u^3-{\ds\frac59}v^2 u
+{\ds\frac{20}{9}}u^2v-u_x^2+{\ds\frac59}v_x u_x-{\ds\frac{2}{27}}v_x^2.
$$

As a final remark, notice that if we define nonlocal variables by means of the formulas
$w,y,z$
$$
\ba{ll}
w_x = y, & w_t = -(2/3)w v_x+3 w u_x+(4/3) y v-6 y u, \\
y_x = w v/3, & y_t = (2/3) y v_x-3 y u_x+(4/9)w v^2 
-2 w v u-(2/3) w v_{xx}+3 w u_{xx},\\
z_x = 1/w^2, & z_t = -(2/3)(-2 v+9 u)/w^2, \ea
$$
then using the results from \cite{marv2} we can
rewrite $\mathfrak{R}$ in the standard form, with
$D^{-1}$ appearing in each term at most once:
$$
\ba{l} \mathfrak{R}=\left( \ba{cc}
3 & -2/3\\
6 & -1%
\ea\right) D^2+\left( \ba{cc}
- 6u  & v/9+2u/3\\
6 u - 9v & 8 v/3- 6u
\ea\right)
+ \sum\limits_{\alpha=1}^4\boldsymbol{K}_\alpha
D^{-1}\circ\boldsymbol{\gamma}_\alpha. \ea
$$
Here $\boldsymbol{\gamma}_1=(-3 , 1/2)$,
$\boldsymbol{\gamma}_\alpha=(0 , w^2 z^{\alpha-2})$,
$\alpha=2,3,4$, are nonlocal cosymmetries for (\ref{eveq}),
$\boldsymbol{K}_1=(u_x,  v_x)^T$ is a local symmetry of  (\ref{eveq}),
and $\boldsymbol{K}_\alpha$, $\alpha=2,3,4$, are {\em nonlocal}
symmetries of (\ref{eveq}) of the form \looseness=-1
$$
\ba{l} \boldsymbol{K}_2=(-\frac{1}{27} (9 u_{xxx} - 3
v_{xxx}
- 27 u u_x + 21 v u_x 
 + 12 u v_x - 5 v v_x) w^2 z^2
+ \frac{1}{27} (-54 u_{xx} + 15 v_{xx}\\[2mm]
+ 36 u^2 - 30 u v + 4
v^2) w y z^2
+ \frac{1}{54} (-108 u_{xx} + 30 v_{xx} 
 - 99 y^2 z u_x + 24 y^2 z v_x + 72 u^2 - 60 u v + 8 v^2) z \\[2mm]
+ \frac{1}{9} z y(-33 u_x + 8 v_x)/w + (-\frac{11}{6} u_x + \frac{4}{9} v_x)w^{-2}, 
-\frac{1}{9} (9 u_{xxx} - 3 v_{xxx}- 18 (u - v) u_x + 9 u v_x - 4 v v_x) w^2 z^2 \\[2mm]
+ \frac{1}{9} (-63 u_{xx} + 18 v_{xx} + 36 u^2 - 36 u v + 5 v^2) w y z^2 
+ \frac{1}{18} (36 v_{xx}-126 u_{xx}
+ y^2 z (39 v_x - 162 u_x) \\[2mm]
+ 72 (u^2 - u v) + 10 v^2) z
+ \frac{1}{3} z y(-54 u_x + 13 v_x)/w 
+ (-9 u_x + \frac{13}{6} v_x)w^{-2})^T, \ea
$$
$$
\ba{l}
\boldsymbol{K}_3=(\frac{2}{27} (9 u_{xxx} - 3 v_{xxx} - 27 u u_x 
+ 21 v u_x + 12 u v_x - 5 v v_x) w^2 z
+ 2 u_{xx} - \frac{5}{9} v_{xx}  \\[2mm]
- \frac{2}{27} (-54 u_{xx} + 15 v_{xx} + 36 u^2 - 30 u v + 4 v^2) w y z 
+ (y^2 z+y/w)(\frac{11}{3} u_x - \frac{8}{9} v_x)
 - \frac{4}{3} u^2+ \frac{10}{9} u v - \frac{4}{27} v^2, \\[2mm]
\frac{2}{9} (9 u_{xxx} - 3 v_{xxx}
 - 18 (u-v) u_x  + 9 u v_x - 4 v v_x) w^2 z 
+ 7 u_{xx} - 2 v_{xx} - \frac{2}{9} (-63 u_{xx} + 18 v_{xx}\\[2mm]
 + 36 u^2 - 36 u v + 5 v^2) w y z
+ 18 y^2 z u_x - \frac{13}{3} y^2 z v_x 
- \frac{1}{3} (-54 u_x + 13 v_x) y/w
 - 4 u^2 + 4 u v - \frac{5}{9} v^2)^T,
\ea
$$
$$
\ba{l}
\boldsymbol{K}_4=((-\frac{1}{3} u_{xxx}
+ \frac{1}{9} v_{xxx} + u u_x - \frac{7}{9} v u_x - \frac{4}{9} u v_x 
+ \frac{5}{27} v v_x) w^2 + \frac{1}{27} (-54 u_{xx} + 15 v_{xx}\\[2mm]
+ 36 u^2 - 30 u v + 4 v^2) w y
+ \frac{1}{18} (-33 u_x + 8 v_x) y^2, 
(-u_{xxx} + \frac{1}{3} v_{xxx} + 2 u u_x - 2 v u_x - u v_x + \frac{4}{9} v v_x) w^2 \\[2mm]
+ \frac{1}{9} (-63 u_{xx} + 18 v_{xx} + 36 u^2 - 36 u v + 5 v^2) w y 
+ \frac{1}{6} (-54 u_x + 13 v_x) y^2)^T. \ea
$$

It would be interesting to investigate the properties of nonlocal
symmetries $\boldsymbol{Q}_{\alpha,j}\equiv
\mathfrak{R}^j(\boldsymbol{K}_\alpha)$, $j=1,2,\dots$, $\alpha=2,3,4$,
and in particular to find out whether the commutators of $\boldsymbol{Q}_{\alpha,j}$
with local symmetries $\boldsymbol{Q}_k$
yield any new symmetries for (\ref{eveq}). We intend to address
these and related issues elsewhere.

\subsubsection*{Acknowledgments}
This research was supported in part by the Czech Grant Agency 
under grant No. 201/04/0538 and
the Ministry of Education, Youth and Sports of Czech Republic under grant MSM:J10/98:192400002.


\end{document}